\documentclass[12pt]{article}

\newcommand{\be}{\begin{equation}}
\newcommand{\ee}{\end{equation}}
\newcommand{\beqs}{\begin{eqnarray}}
\newcommand{\eeqs}{\end{eqnarray}}
\newcommand{\LL}{{\cal L}}

\newcommand{\half}{{1 \over 2}}

\newcommand{\thab}{{\theta^{\alpha \beta}}}
\newcommand{\thmn}{{\theta^{\mu \nu}}}
\newcommand{\st}{{\vartheta}}
\newcommand{\dxy}{{\frac{dx dy d\sigma}{\sqrt \theta}\,}}

\newcommand{\dy}{{\frac{dy d\sigma}{\sqrt {\theta(X)}}\,}}

\newcommand{\dxR}{{\frac{dx d\phi}{\sqrt \theta}\,}}
\newcommand{\thRi}{{\theta^{0i} (R)}}
\newcommand{\thRj}{{\theta^{0j} (R)}}
\newcommand{\thiR}{{\theta^{i0} (R)}}
\newcommand{\thij}{{\theta^{ij}}}

\newcommand{\thijo}{{\theta_o^{ij}}}
\newcommand{\thjio}{{\theta_o^{ji}}}
\newcommand{\thRio}{{\theta_o^{0i}}}

\expandafter\ifx\csname mathbbm\endcsname\relax

\else

\fi
\begin{document}
\begin{titlepage}
\begin{flushleft}  
       \hfill                       CCNY-HEP-04/8\\
       \hfill                       August 2004\\
\end{flushleft}
\vspace*{3mm}
\begin{center}
{\LARGE {Chiral actions from phase space (quantum Hall) droplets}\\}
\vspace*{12mm}
\large Alexios P. Polychronakos \\
\vspace*{5mm}
\large
{\em Physics Department, City College of the CUNY\\
Convent Avenue and 138$^{th}$ Street, New York, NY 10031\\
\small alexios@sci.ccny.cuny.edu \/}\\
\vspace*{4mm}
\vspace*{15mm}
\end{center}
%\maketitle

\begin{abstract} {
We derive the hamiltonian and canonical structure for arbitrary
deformations of a phase space (quantum Hall) droplet on a general
manifold of any dimension. The derivation is based on a transformation
that decouples the Casimirs of the density Poisson structure.
The linearized theory reproduces the edge state chiral
action of the droplets, while the nonlinear hamiltonian captures
$1/N$ quantum corrections.}

\end{abstract}

\end{titlepage}

\section{Introduction}

Collective states of fermions have nontrivial dynamics and are at the
source of many interesting physical phenomena. Effective descriptions
of such states in terms of bosonic fields that capture their essential
properties are, therefore, very desirable. Such descriptions are mostly
available in one spatial dimension and particular reductions of
higher dimensional systems \cite{Col}-\cite{Pol}.

Although fermion interactions are in general an important ingredient,
there are situations in which the fermions are effectively noninteracting,
the nontrivial behavior of their collective state arising from the
structure of their phase space and Fermi exclusion. In such situations
a fluid dynamical description of the system in terms of its phase space
density becomes particularly well-suited \cite{Pol}. The typical example
of this is the integer quantum Hall effect, where incompressibility and
edge excitations can be studied this way \cite{WeLe}-\cite{CaDuTruZe}.
(Fractional quantum Hall states are also amenable to a similar
description but with generalized phase space statistics.)

Generalizations of the quantum Hall system to higher dimensions have been
proposed by Zhang and Hu and led to some interesting speculations 
\cite{ZhaHu}-\cite{BeCaNe}. This construction has been extended to various
geometries, topologies and background gauge fields, and the properties 
of these theories have been extensively studied. In particular, Karabali
and Nair gave a general method for deriving an effective action for the 
edge excitations of quantum Hall droplets \cite{KaNa}. This
action is effectively chiral and, in the nonabelian case, constitutes
an interesting higher dimensional generalization of the chiral sigma model
known as the Wess-Zumino-Witten model \cite{WZW}. 

The derivation of this effective action was based on the quantum
mechanical density matrix formulation and a large-$N$ approximation
(small perturbation of the droplet) \cite{Sak}. The dynamics of the quantum
Hall droplet, on the other hand, are captured in an essential way by the
classical fluid motion in the phase space of the droplet.
The purpose of the present paper is to present a derivation of the
chiral droplet action in a general setting, without the assumption of
small perturbation, by analyzing their classical phase space.
A nonlinear theory is obtained that describes arbitrary deformations
of the droplet. Quantum mechanically, the nonlinear terms capture
higher order corrections in the $1/N$ approximation. The analysis is
presented for the abelian case (spinless, colorless fermions), deferring
the treatment of nonabelian droplets to a future publication.

The organization of the paper is as follows. In section 2 we give a
general analysis of phase space density dynamics and introduce the
cartographic transformation, which renders the canonical structure of
the density theory amenable to a lagrangian realization.
In section 3 we analyze the motion of droplets and apply the formalism
of section 2 to derive their hamiltonian and Poisson structure. A
linearized analysis is performed, recovering the chiral action of
Karabali and Nair. In section 4 we examine the quantization of this
theory in 2 dimensions and demonstrate that the expected results
are obtained including corrections of order $1/N$. Finally, section
5 contains our conclusions and discussion of outstanding issues.

\section{Phase space density dynamics}
\subsection{General formulation}

We shall consider noninteracting particles moving in a general
$D$-dimensional phase space with an arbitrary hamiltonian ($D$ must be
even for a non-degenerate canonical structure). Although
non-relativistic particles will be relevant for quantum Hall considerations,
in principle the motion could be relativistic, depending on the structure
and symmetries of the hamiltonian and phase space. External electric and
magnetic fields are also included and encoded in the canonical structure.

The single-particle phase space will be described by phase space coordinates
$\phi^\alpha$, $\alpha = 0,1,\dots D-1$ and determined by its Poisson
structure:
\be
\{ \phi^\alpha , \phi^\beta \}_{sp} = \thab
\label{spPB}
\ee
where, to avoid later confusion, we put the subscript $sp$ standing 
for single-particle.

The Poisson matrix $\thab$ will be, in general, a function
of the phase space coordinates $\phi^\alpha$. The volume element in this
phase space is, then,
\be
dv = \frac{d\phi}{\sqrt{\theta}} ~~,~{\rm where}~~ \theta = \det\thab
~,~~ d\phi = \prod_{\alpha=0}^{D-1} d \phi^\alpha
\ee
The single-particle hamiltonian, denoted $V(\phi)$, leads to classical
motion:
\be
{\dot \phi}^\alpha = \{ \phi^\alpha, V \}_{sp} = \thab \partial_\beta V
\ee

A dense collection of particles on this phase space can be described in terms
of its density $\rho (\phi,t)$. Under time evolution, $\rho$ changes by 
a canonical transformation generated by $V$:
\be
{\dot \rho} = \{ \rho , V \}_{sp} = \thab \partial_\alpha \rho
\partial_\beta V
\label{eomrho}
\ee
The above equation can arise out of a hamiltonian and canonical structure
for the field $\rho$. Choosing for the hamiltonian the total particle
energy
\be
H = \int \frac{d\phi}{\sqrt{\theta}} \rho V
\ee
the appropriate Poisson brackets are
\be
\{ \rho(\phi_1 ) , \rho(\phi_2 ) \} = \sqrt{\theta(\phi_+ ) }
\thab (\phi_+ ) \partial_\alpha \rho (\phi_+ ) \partial_\beta \delta
(\phi_- ) 
\label{P}
\ee
where we defined relative and mid-point coordinates $\phi_- = \phi_1
- \phi_2$ and $\phi_+ = \frac{\phi _1 + \phi_2}{2}$. 
[These brackets should not
be confused with the single-particle brackets (\ref{spPB}).] 
It can be checked that the canonical equation of motion
for the field $\rho$:
\be
{\dot \rho} = \{ \rho , H \}
\ee
reproduces equation (\ref{eomrho}). In the derivation we need to use
the identity
\be
\partial_\alpha \left( \frac{\thab}{\sqrt{\theta}} \right) = 0
\label{ident}
\ee
which is a corollary of the Jacobi identity for $\thab$.

The above brackets (\ref{P}) are the standard infinite-dimensional
Poisson algebra on the phase space manifold. In terms of test functions
their form becomes more obvious: defining
\be
\rho[F] = \int \frac{d\phi}{\sqrt{\theta}} F(\phi) \rho(\phi)
\ee
for some function on the phase space $F$, then the brackets of two 
such integrals are
\be
\{ \rho[F] , \rho [G] \} = \rho[ \{ F, G \}_{sp} ]
\label{PB}
\ee

A lagrangian realization  of the equation of motion for the field $\rho$
in terms of an action would require the inversion
of the Poisson structure (\ref{P}). This, however, is obstructed by the
fact that the above Poisson brackets are degenerate; that is, the above
algebra has Casimirs. Indeed, for any function of a single variable $f(x)$,
the integral
\be
C[f] = \int \frac{d\phi}{\sqrt{\theta}} f(\rho) 
\ee
has vanishing Poisson brackets with $\rho$ and constitutes a Casimir. There
are, thus, an infinite number of Casimirs. They are spanned, e.g.,
by $C_n \equiv C[x^{n+1} ]$ for $n=0,1,2,\dots$.

A lagrangian realization, then, of the above structure could proceed
in two possible ways: we could either neutralize the Casimirs by fixing
them to some numerical values, thus reducing the phase space, or append
canonical partners to the Casimirs to render the Poisson brackets
non-degenerate, thus augmenting the phase space. The first option,
for instance, could be implemented by considering only density fields
$\rho$ that are generated by canonical transformations on the single-particle
phase space of some reference density function $\rho_0$. This is the
approach that would most closely parallel the treatment in quantum
mechanics. 

In the next section we shall present an alternative way to realize
the above structure that helps decouple the Casimirs, by transforming
to a new set of dynamical variables.
The transformation is motivated by the droplet picture and its significance
will manifest when we apply it to the canonical formulation of droplets.

\subsection{The cartographic transformation}

To derive the desired transformation it is conceptually and analytically
simpler to start from a more general framework, and later specialize to
the problem at hand. 

Consider a (new) classical phase space of dimension $D+1$ (which could
be either even or odd). For later
convenience, we single out in notation two of the coordinates of this
space, $x$ and $y$, while we call the remaining coordinates $\sigma^i$,
$i=1,2,\dots D-1$. 
The canonical structure of the manifold is $\thmn (x,y,\sigma)$.
(In the sequel we use middle greek letters for the full set
of indices $(x,y,1,2, \dots D-1)$ and latin letters for the indices of $\sigma$,
while early greek letters will be reserved for indices taking values in
a $D$-dimensional space.)

The dynamical variable will be a function of one less
variable, which we take to be $x$: $X (y,\sigma )$. The field $X$ will 
possess Poisson brackets to be defined shortly.

We can perform a transformation of the field $X$ into a new field $Y$
by trading one of the remaining independent variables, take it to be $y$,
with $X$ itself. Specifically, we define  the field $Y(x,\sigma)$
such that
\be
X(Y(x,\sigma),\sigma) = x ~,~~~ Y(X(y,\sigma),\sigma) = y
\ee
To give an explicit expression for this transformation, define the step
function $\vartheta (x)$ as
\be
\st (x) = \half [(1+sgn(x)] = 1 ~{\rm if}~x>0~,~~ =0 ~{\rm else} 
\ee
Assuming proper monotonicity properties for $X$ we have
\be
\st (X-x) = \st (Y-y)
\label{thth}
\ee
[while, in general, $f(X-x) \neq f(Y-y)$]. Using the above and the
fact that $\vartheta' (x) = \delta (x)$, the transformation becomes
\be
Y(x,\sigma ) = \int dy \, y \delta (Y-y) = \int dy  \st (Y-y) 
= \int dy \st (X-x)
\label{hyps}
\ee
while the inverse transformation is
\be
X(y,\sigma ) = \int dx \st (Y-y)
\ee
This transformation is obviously symmetric.

It is useful to visualize the fundamental dynamical variable as a
$D$-dimensional membrane (a ``$D$-brane'')
embedded in the $D+1$-dimensional phase space. $X$ and $Y$,
then, are two different ways of parametrizing this membrane in terms
of its $x$ or $y$ coordinate, respectively. The transition from
$X$ to $Y$ will be called a cartographic transformation.
When we specialize to the case of interest, namely
of a phase space density $\rho$ in $D$ phase space dimensions, $X$
will become the density, while $Y$ will effectively become
a parametrization of the density in terms of overlapping droplets.
This shall be explained in the subsequent sections.

To posit the fundamental Poisson brackets of the problem, define the
quantity $S[A]$ for an arbitrary function $A(x,y,\sigma)$:
\be
S[A] = \int \dxy A(x,y,\sigma) \st (X-x) = \int \dxy A(x,y,\sigma) \st (Y-y)
\label{SA}
\ee
We postulate the Poisson brackets of the above dynamical variables as
\be
\{ S[A] , S[B] \} = \int \dxy \{ A, B \}_{sp} \st (X-x) 
= \int \dxy \thmn \partial_\mu A \, \partial_\nu B \, \st (X-x)
\label{PBAB}
\ee
where $\{ A, B \}_{sp}$ are the single-particle Poisson brackets
in the full phase space.
The Poisson brackets (\ref{PBAB}) satisfy the Jacobi identity,
as a corollary of the Jacobi identity for $\thmn$.
By using (\ref{thth}) we can obtain equivalent expressions for the $S[A]$
and their Poison brackets in terms of $Y(x,\sigma)$.

The above Poisson brackets imply underlying Poisson brackets for the
fundamental dynamical fields $X$ or $Y$. To determine them for $X$, 
substitute the following specific functions in the definition of 
$S[A]$ and $S[B]$
\be
A = {\sqrt \theta} \delta (y-y_1 ) \delta (\sigma - \sigma_1 ) ~,~~~
B = {\sqrt \theta} \delta (y-y_2 ) \delta (\sigma - \sigma_2 )
\label{AB}
\ee
which gives $S[A] = X(y_1 , \sigma_1 )$, $S[B] = X(y_2 , \sigma_2 )$.
Plugging the above functions in (\ref{PBAB}) we obtain after some algebra
\be
\{ X (y_1 , \sigma_1 ) , X(y_2 , \sigma_2 ) \} = \sqrt{\theta_+ (X)}
\left[ \theta_+^{x \alpha} (X) 
+ \theta_+^{\alpha \beta} (X) (\partial_\beta X)_+ \right]
\partial_\alpha \delta (y_- , \sigma_- ) 
\label{PBX}
\ee
where the indices $\alpha,\beta$ take the values $y,i$. In the above,
we defined
\be
\theta_+^{\mu \nu} (X) = \theta^{\mu \nu} (X(y_+ , \sigma_+ ), y_+ , \sigma_+ )
~,~~~ (\partial_i X)_+ = (\partial_i X) (y_+, \sigma_+ )
\ee

Conversely, assuming the Poisson brackets (\ref{PBX}) for $X$ we can show
that (\ref{PBAB}) are satisfied. (This step is necessary, since the set
of functions (\ref{AB}) used to derive (\ref{PBX}) do not span the full
space of functions of three variables.) The derivation is given below.

To alleviate the notation, for any function $f(x,y,\sigma )$ (such as
$A$ or $\theta^{\mu \nu}$) we will drop the arguments $x$, $y$ and $\sigma$
we will write
\be
f(X) \equiv f(X(y,\sigma ),y,\sigma ) ~,~~~ 
(\partial_\alpha f)(X) \equiv \partial_\alpha f(x,y,\sigma )|_{x=X}
\ee
while $\partial_\alpha f(X)$ will stand for the total derivative of $f(X)$,
both explicit and implicit through $X(y,\sigma )$.
(As before, early greek indices stand for $y$ or $i$.)

The Poisson brackets of two functionals $F$ and $G$ of $X$, will be given
in terms of their variations with respect to $X$ and the Poisson brackets 
of $X$ (\ref{PBX}). Writing
\be
\frac{\delta F}{\delta X} = \frac{1}{\sqrt {\theta (X)}} f ~,~~~ 
\frac{\delta G}{\delta X} = \frac{1}{\sqrt {\theta (X)}} g  
\ee
and using (\ref{PBX}) we have
\beqs
\{ F , G \} &=& \int dy_1 d\sigma_1 dy_2 d\sigma_2
\frac{\delta F}{\delta X_1} \frac{\delta G}{\delta X_2} \{ X_1 , X_2 \} \cr
&=& \int \dy f \left[ \theta^{x\alpha} (X) + \theta^{\alpha \beta} (X) 
\partial_\beta X \right] \partial_\alpha g
\eeqs
We shall apply this formula for the functionals $S[A]$ in (\ref{SA}),
whose variation is
\be
\delta S[A] = \int \dxy A \, \delta (X - x) \, \delta X =
\int \dy A(X) \, \delta X
\ee
We obtain
\be
\{ S[A] , S[B] \} = \int \dy A(X) \, \left[ \theta^{x\alpha} (X) + 
\theta^{\alpha \beta} (X) \partial_\beta X \right] \partial_\alpha B(X)
\ee
Expanding the total derivative $\partial_\alpha B(X) = (\partial_\alpha B)(X)
+ (\partial_x B)(X) \, \partial_\alpha X$ we have
\beqs
\{ S[A] , S[B] \} = \int \dy A(X) &\Bigl[&\theta^{x\alpha} (X) \,
(\partial_\alpha B)(X) + \theta^{x\alpha} (X) \,
(\partial_x B)(X) \, \partial_\alpha X  \cr
&+& \theta^{\alpha \beta} (X) \, (\partial_\alpha B)(X) \,
\partial_\beta X ~\Bigr]
\eeqs
By inserting $\int dx \delta (X-x)$ in the above integral we can eliminate
the dependence of all quantities on $X$ and reintroduce $x$. Further, by using
\be
\delta (X-x) = -\partial_x \st (X-x) ~,~~~
\partial_\alpha X \delta (X-x) = \partial_\alpha \st (X-x)
\ee
the above expression becomes
\beqs
\{ S[A] , S[B] \} = \int \dxy A &\Bigl[& \theta^{\alpha x} \, 
\partial_\alpha B \, \partial_x \st (X-x) 
+ \theta^{x\alpha} \, \partial_x B \, \partial_\alpha \st (X-x) \cr
&+& \theta^{\alpha \beta} \, \partial_\alpha B \, \partial_\beta \st (X-x) 
~ \Bigr]
\eeqs
We see that the terms in the bracket constitute the full expansion
of the single-particle bracket $\{ B , \st (X-x) \}_{sp}$.
With an extra integration by parts, and using $\partial_\mu ( \thmn
/ \sqrt{\theta} ) = 0$, we obtain the Poisson bracket (\ref{PBAB})
as desired.

A similar analysis would yield the Poisson brackets for $Y$. The result is
\be
\{ Y (x_1 , \sigma_1 ) , Y (x_2 , \sigma_2 ) \} = \sqrt{\theta_+ (Y)}
\left[ \theta_+^{y \alpha} (Y) 
+ \theta_+^{\alpha \beta} (Y) (\partial_\beta Y)_+ \right] \partial_\alpha
\delta (x_- , \sigma_- )
\label{PBY}
\ee
in an obvious notation, where now $\alpha,\beta=x,i$.

To recapitulate, we have demonstrated that if the variable $X(y,\sigma)$
has Poisson brackets (\ref{PBX}), then the transformation (\ref{hyps})
generates a new variable $Y(x,\sigma)$ that has Poisson brackets
(\ref{PBY}).

If the canonical structure $\thmn$ is non-degenerate, the Poisson structure
of $X$ or $Y$ will also be non-degenerate, the only Casimir being the
trivial one corresponding to $A=1$:
\be
C_0 = \int \dxy \st (X-x) = \int \dxy \st (Y-y)
\ee
The rest of the Casimirs $C_n$ constructed in analogy to the ones for $\rho$ 
are not present here (or, rather, they are all equal to $C_0$),
since $\st (x)^{n+1} = \st (x)$ for $n=0,1,2,\dots$. Note, further,
that $S[A]$ span all functionals of $X$ or $Y$ of
the form $\int f(X,y,\sigma)$ or $\int f(x,Y,\sigma)$.
Indeed, any function $f(X)$ can be expanded in terms of step functions as
\be
f(X) = \int dx f' (x) \st (X-x)
\ee
and with functions $A(x,y,\sigma)$ this can be done pointwise on the
space $(y,\sigma)$. Therefore, the algebra of all these functionals
is non-degenerate and contains no Casimirs except $C_0$. This non-degeneracy
in the Poisson brackets of $X$ or $Y$ manifests in the appearance of
the affine term proportional to $\theta^{x\alpha}$ or $\theta^{y\alpha}$.
If, on the other hand, the canonical structure $\thmn$ is degenerate,
this degeneracy will also be inherited by the Poisson structures of
$X$ or $Y$.

\subsection{Canonical decomposition of the density}

We are now ready to apply this formalism to the case of the density field 
$\rho$ on our original phase space. For this, choose the $D+1$-dimensional
phase space to be our original $D$-dimensional phase space plus one
additional trivial coordinate $x$. That is, choose $\thmn (y,\sigma)$
independent of $x$ given by
\be
y = \phi^0 ~,~~ \sigma^i = \phi^i ~;~~~
\theta^{x\alpha} = 0 ~,~~
\thab (y,\sigma) = \thab (\phi^0 , \phi^i)
\ee
The function $X(y,\sigma)$ will be identified with the phase space
density $\rho$. With the above choice it is clear that the affine
term in (\ref{PBX}) as well as all dependence on $X=\rho$ drop and we recover
the standard Poisson brackets of $\rho$ (\ref{P}).

This canonical
structure is certainly degenerate, due to the degeneracy of the coordinate
$x$, making all functions of $x$ Casimirs of the single-particle Poisson
brackets. As a consequence, the Poisson structure of $X=\rho$ is also
degenerate. Indeed, the set of $S[f(x)]$ for any function of $x$ alone 
constitute Casimirs, spanned by $S[x^n]$ for $n=0,1,\dots$
Substituting $A=x^n$ in (\ref{SA}) and using the fact that $\theta$ is
$x$-independent and $X=\rho$ we obtain
\be
S[x^n] = \frac{1}{n+1} \int \frac{d\phi}{\sqrt{\theta}} \rho^{n+1}
\ee
recovering the tower of Casimirs $C_n$ identified before.

Perform, now, the cartographic transformation from $x$ to $y=\phi^0$,
The hamiltonian and Poisson structure for $Y$ become
\be
H = \int \frac{dx d\phi}{\sqrt \theta} V(\phi) \st (Y-\phi^0 )
\ee
and
\be
\{ Y (x_1 , \sigma_1 ) , Y (x_2 , \sigma_2 ) \} = \sqrt{\theta_+ (Y)}
\left[ \theta_+^{0 \alpha} (Y) 
+ \theta_+^{\alpha \beta} (Y) (\partial_\beta Y)_+ \right] \partial_\alpha
\delta (x_- , \sigma_- )
\label{PBYR}
\ee
Note that the above Poisson brackets do not involve any $x$-derivatives.
As a result, the Poisson algebra decomposes into an infinite direct sum
of mutually commuting algebras, one for each value of $x$. Each of these
algebras is non-degenerate (note that (\ref{PBYR}) still contains
an affine term), the only Casimir being the zero-mode of $Y$:
\be
C (x) = \int \frac{d\phi}{\sqrt \theta} \st (Y - \phi^0 )
\ee
The above reproduces the full set of Casimirs
of the original Poisson algebra. Indeed, using 
$\st (Y - \phi^0 ) = \st (\rho - x)$ and
the $x$-independence of $\theta$ we have
\be
\int dx x^n C(x) = \int \dxR x^n \st (\rho -x) = \frac{1}{n+1}
\int \frac{d\phi}{\sqrt \theta} \rho^{n+1} = \frac{1}{n+1} C_n
\ee

We therefore achieved to bring the density Poisson algebra in a 
decoupled form that makes its lagrangian realization possible:
For fixed $x$, the algebra of $Y(x)$ can be realized in terms of an
action, the only obstruction to take care being the zero-mode $C(x)$.
This can be done either by omitting the zero mode altogether (fix
the Casimir $C(x)$) or by endowing it with a canonical partner $\Theta (x)$.
Specifically, if we add an extra field $\Theta (x)$ (a function of
$x$ only) with Poisson brackets
\be
\{ \Theta (x_1 ) , Y(x_2 , \sigma ) \} = \sqrt{\theta (Y)} \, \delta (x_- )
\ee
then the full algebra becomes non-degenerate.

\subsection{Comparison to fluid dynamics}

It is useful to note that the above cartographic transformation is similar
to the hodographic transformation that produces a fluid mechanical system
out of a Nambu-Goto $D$-brane theory \cite{BoHo,JaPo}.
(For a comprehensive review of this and related issues see \cite{JNPP}.)
The differences are that in our case we work in phase space
while in the $D$-brane case there are independent momenta, and
that in our case we only trade one variable while in the $D$-brane case
$D$ spatial world-volume variables are traded for corresponding
target-space variables (hodographic, which means `path-tracing' in
Greek, alludes to this point-wise mapping in time from target space
to world volume space). 

In the $D$-brane case this mapping has the
effect of reverting from a `lagrangian' (body-fixed) description of
the fluid to an `eulerian' one in terms of density and velocity fields.
The volume-preserving reparametrization invariance of the former
description (a `particle re-labeling' transformation) is eliminated
by working in terms of the reparametrization-invariant quantities
of density and velocity.

In our case, something similar takes place:
an underlying particle realization of the phase space density $\rho$
would introduce particle-relabeling arbitrariness, which is related to
the existence of the Casimirs $C_n$.
The cartographic transformation is more like a transition from an
eulerian description in terms of a phase-space density $\rho$ (there are no
velocities) to a partially lagrangian description in terms of $Y$. That
$Y$ is akin to a lagrangian body-fixed variable is evident from its
equation of motion as will be derived in the next section, which contains
a term relevant to a particle at position $\phi^0 = Y$. This description,
however, neutralizes the Casimirs by eliminating the relabeling arbitrariness
of particles along the contours $\rho$=constant. It amounts to parametrizing
the landcape of the field $\rho$ in terms of its curves of equal altitude,
thus the name `cartographic'.

We conclude with the comment that the above transformation resolves
in principle the problem of the lagrangian realization of the $\rho$-algebra,
although in practice the explicit realization of the algebra of $Y(x)$
will depend on the exact form of $\thab ( \phi )$ and may be complicated.

\section{Phase space droplet dynamics}
\subsection{Description and equation of motion}

We now turn our attention to the situation where the particles underlying
the density $\rho$ are (spinless) fermions.
A dense collection of fermions in the phase space will form a 
Fermi liquid. Semiclassically, the fermions will fill densely a region of the
phase space, with one particle per volume $h^{D/2}$. Such a state constitutes 
a constant density droplet of arbitrary shape, uniquely determined
by its boundary. Under time evolution, the density remains constant
inside the droplet (by Liouville's theorem) while each point of the
boundary moves according to the single-particle equation of motion, thus
deforming the shape of the droplet.

To describe the droplet it suffices to determine the shape of its
$D-1$-dimensional boundary. This can be done by expressing one of the
phase space coordinates on the boundary, say $\phi^0 \equiv R$, as a 
function of the remaining phase space coordinates $\sigma^i$. So the dynamical
variable is the function $R(\sigma,t)$. (For a finite droplet, it is
convenient to assume that the origin of coordinates is inside the droplet
and to think of $R$ as a `radial' coordinate and $\sigma^i$ as `angular'
coordinates.)

The evolution of $R$ is due to two factors: first, the coordinate 
$R$ itself evolves according to the single-particle equation of motion; 
and second, the variables $\sigma^i$ evolve as well. A point on the
boundary evolves as
\be
(R,\sigma) \to (R + {\dot \phi}^0 |_{\phi^0 = R} dt, \sigma + {\dot \sigma}
dt )
\ee
The canonical evolution of $\phi^0$ at the boundary, then, gives the
co-moving time derivative of the function $R(\sigma )$:
\be
{\dot R} + {\dot \sigma}^i \partial_i R = \thRi \, (\partial_i V)(R)
\ee
Expressing ${\dot \sigma}^i$ in terms of its equation of motion
gives the time derivative of $R$:
\be
{\dot R} = \thRi \, (\partial_i V)(R) 
-\thiR \, (\partial_0 V)(R) \, \partial_i R
-\thij (R)  \, (\partial_j V)(R) \, \partial_i R
\ee
The total $\sigma$-derivative of the hamiltonian function on the boundary 
$V(R)$ is
\be
\partial_i V(R) = (\partial_i V)(R) + (\partial_0 V)(R) \, \partial_i R
\ee
Using the above, as well as the antisymmetry of $\thij$, we can express
the equation of motion for $R$ in terms of the total derivative of
$V(R)$ as:
\be
{\dot R} = \thRj \,\partial_j V(R) -\thij (R) \,\partial_j V(R) \,\partial_i R
\label{eomR}
\ee
This is the basic equation of motion for the droplet. It is first-order in
time and phase space derivatives and thus fundamentally chiral.

It is worth noting
that, if the coordinate $\phi^0$ is chosen to parametrize the potential
(that is, surfaces $\phi^0$=constant are equipotential), $V=V(\phi^0 )$,
the second term above drops and we get
\be
{\dot R} = \thRi \partial_i V(R) 
\ee
In the special case when $\thRi$ is nonzero only for a single value of the
index $i$ (there is a global variable conjugate to $\phi^0$, call it $\phi^1$),
and is a function only of $\phi^0$, the above equation becomes
\be
{\dot R} = \theta^{01} (R) V' (R) \partial_1 R
\ee
which can easily be solved by a hodographic transformation, interchanging
$R$ and $\phi^1$. 
Such choices will be useful later, but for now we remain fully general.

We should warn that depending on its topology the droplet may have more than
one boundaries. In such cases we would need to introduce several commuting
boundary fields $R_n$, one for each boundary. Similarly, the boundary could
intersect $\sigma$=constant lines at more than one $\phi^0$, in which
case we would again need to introduce several boundary fields, one
for each branch, with appropriate matching conditions tying them into
a unique boundary.

\subsection{Canonical realization of droplet dynamics}

The above dynamics arise from a hamiltonian reduction
of the full density canonical structure: a constant-density droplet of
arbitrary shape constitutes a particular class of density functions and
thus a submanifold of the full manifold of configurations for $\rho$,
of the form
\be
\rho = \rho_o \st (R-\phi^0 )
\ee
We need to project the canonical two-form of $\rho$ on this submanifold.
This can be easiest done in the cartographic transformed variable. Applying
(\ref{hyps}) for this $\rho$, and noting that
\be
\st ( \rho_o \st (R-\phi^0 ) -x) = \st (R-\phi^0 ) \st (\rho_o - x)
\ee
we obtain
\be
Y(x,\sigma) = \st (\rho_o - x) R(\sigma)
\ee
So the reduction consists in putting all $Y(x)=0$ for $x>\rho_o$
and all $Y(x)$ equal to each other for $x<\rho_o$. This second
class constraint is easily implemented, giving a canonical two-form
for $R$
\be
\Omega_R = \int dx \st(\rho_o -x) \, \Omega_{Y(x)} = \rho_o \, \Omega_Y
\ee
where we used the decoupling in $x$ and $x$-independence of the
canonical structure of $Y(x)$. Therefore, the Poisson brackets for $R$ are
the same as those of $Y(x)$ times $1/\rho_o$:
\be
\{ R(\sigma_1 ) , R(\sigma_2 ) \} = \frac{\sqrt{\theta_+ (R)}}{\rho_o} 
\left[ \theta_+^{0i} (R) + \theta_+^{ij} (\partial_j R)_+ \right] 
\partial_i \delta (\sigma_- )
\label{PR}
\ee
The corresponding hamiltonian is
\be
H = \int \dxR V(\phi) \st (Y-\phi^0 ) = \rho_o \int 
\frac{d\phi}{\sqrt {\theta}} V(\phi) \st (R-\phi^0 ) 
\ee

The above Poisson structure and hamiltonian encode the full dynamics
of the droplet. We can explicitly verify that they reproduce the equation
of motion (\ref{eomR}) as the canonical evolution
\be
{\dot R} = \{ R , H \}
\ee
Note that the above hamiltonian is defined in the bulk of the droplet,
although the equation of motion refers only to the boundary. 

The constant $\rho_o$ is irrelevant for classical dynamics.
The semiclassical interpretation of the droplet, however, fixes the value
$\rho_o = 1/(2\pi \hbar)^{D/2}$, which will be important for quantization.

We conclude this section with the following remarks:

1. The droplet situation closely parallels the one discussed in
section 3, in dimension $D$ rather than $D+1$: the boundary of the droplet
plays the role of the $D-1$-brane in the phase space parametrized
by its $\phi^0$-coordinate $R$. The hamiltonian is $H=\rho_o S[V]$ 
and the Poisson brackets (\ref{PBAB}) are the statement that the droplet
density satisfies the proper Poisson density algebra (\ref{P}).

2. The Poisson brackets (\ref{PR}) of $R$ contain an affine `chiral' part 
as well as an ordinary Poisson density structure 
(the second term in the bracket).
over the gauge manifold $\{ \sigma^i \}/\phi_0$. 
The quotient arises because $\thij$ is degenerate,
being odd-dimensional, and effectively the variable conjugate to
$\phi^0$ drops out.

3. ({\ref{PR}) satisfies the Jacobi identity, as a corollary of the
Jacobi identity of the full Poisson brackets for $Y$, although its 
direct check is highly nontrivial. In the special case when $\thab$
is independent of $\phi^0$ the affine and linear terms decouple
and individually satisfy the Jacobi identity. In the
generic case, however, both terms are needed to satisfy the identity.

4. The Casimirs of the original density for the droplet become
$C [x] = \st (\rho_o -x) C_0$ or $C_n = C_0$. So they are all neutralized,
the only essential Casimir being the total particle number $C_0 = N$.

\subsection{The linearized action}

Assuming that the droplet deviates only slightly from an equilibrium
configuration, we can analyze its motion as a perturbation
of its equilibrium shape. This will be useful for large
droplets (large number of particles) whose motion involves only
boundary perturbations.

Consider an equilibrium configuration in which the droplet fills the
phase space up to an energy level $V_o$. That is, the boundary function
$R_o (\sigma)$ is such that
\be
V(R_o (\sigma) , \sigma) = V_o = {\rm constant}
\ee
Such a function obviously satisfies the equation of motion for $R$
with ${\dot R} = 0$. A perturbation of the droplet around this configuration
can be written as $R = R_o + \chi$, $\chi \ll R_o$. Correspondingly, the
energy function is perturbed as 
\be
V(\sigma) = V_o + \omega_o (\sigma) \chi (\sigma) ~~,~~~{\rm where}~~
\omega_o = (\partial_0 V)(R_o )
\ee
So to lowest order in $\chi$ the equation of motion (\ref{eomR})
becomes:
\be
{\dot \chi} = ( \thRio - \thjio \partial_j R_o ) \partial_i 
(\omega_o \chi)
\ee
where $\theta_o^{\alpha \beta} = \thab (R_o (\sigma) , \sigma)$.

Define the differential operator
\be
{\LL} = u^i \partial_i = ( \thRio - \thjio \partial_j R_o ) \partial_i
\ee
Then the above equation becomes simply
\be
{\dot \chi} = \LL ( \omega_o \chi)
\label{eomlin}
\ee
Similarly, to lowest order in $\chi$ we have:
\be
\{ \chi (\sigma_1 ) , \chi (\sigma_2 ) \} = 
\frac{\sqrt{\theta_o (\sigma_+ )}}{\rho_o} \,
\LL (\sigma_+ ) \delta (\sigma_1 - \sigma_2 )
\label{linP}
\ee
and
\be
H = H_o + \rho_o V_o \int \frac{d\sigma}{\sqrt{\theta_o}} \, \chi +
\int \frac{d\sigma}{\sqrt{\theta_o}} \, \half \omega_o \chi^2
\ee
$H_o$ is a constant and can be dropped. The next term, linear in $\chi$,
is simply $\rho_o V_o C_0 $ plus a constant; 
it is therefore a Casimir and does not contribute to the equations of motion. 
It can be set to zero as an initial condition, 
corresponding to a constant-volume perturbation of the droplet
(total number of particles constant), and we end up with 
a quadratic hamiltonian.

We can derive the above linearized structure from a lagrangian and
make contact with the chiral action of Karabali and Nair. To achieve this,
consider that the deformation of the droplet 
is generated by an infinitesimal canonical transformation on the phase space
$\varphi (\sigma)$:
\be
(R_o , \sigma) \to (R_o + \thRio \partial_i \varphi \, , \,
\sigma + \thijo \partial_j \varphi)
\ee
or
\be
\chi = ( \thRio - \thjio \partial_j R_o ) \, \partial_i \varphi
= \LL \varphi
\ee
With this redefinition, the Poisson brackets of $\varphi$ will
involve the inverse of the operator $\LL$ and so its canonical
two-form will involve $\LL$ itself. We obtain the action
\be
S = \int dt \frac{d\sigma}{\sqrt{\theta_o}} \, \half \left[
{\dot \varphi} \LL \varphi - \omega_o (\LL \varphi )^2 \right]
\label{S}
\ee
Note that $\LL$ is an anti-self-adjoin operator on $\sigma$, due to
the relation
\be
\partial_i \left( \frac{u^i}{\sqrt{\theta_o}} \right) = 0
\ee
which is a corollary of (\ref{ident}). The equation of motion from
this action becomes
\be
\LL {\dot \varphi} = \LL ( \omega_o \LL \varphi)
\ee
which, upon putting $\chi = \LL \varphi$, reproduces (\ref{eomlin}).

To make contact with the action of Karabali and Nair, note that in
\cite{KaNa} they effectively made the choice $V=V(R)$ (their
coordinate is `normal' to the boundary) and thus $R_o$=constant.
In this case $\LL = \thRio \partial_i$. They also made the choice
$\omega_o$=constant. With these choices, our action (\ref{S})
becomes identical to theirs.

The zero modes of $\LL$, $\LL \varphi = 0$, do not generate any
transformation for $R$ and should, thus, be discarded. The equation
of motion for $\varphi$, in this case, can be written in the first-order
form:
\be
{\dot \varphi} = \omega_o \LL \varphi
\label{eomfirst}
\ee
On the other hand, there may be topological obstructions to the map
from $\varphi$ to $\chi$, that is, there may be perturbations $\chi$
that cannot be generated by a (single-valued) function $\varphi$.
This is, again, related to zero modes of $\LL$: if $\LL \chi = 0$,
then obvioulsy such a $\chi$ cannot be generated by any $\varphi$.

These zero modes are, in fact, related to Casimirs of the linearized
Poisson brackets (\ref{linP}). Defining $\LL$-eigenfunctions
\be
\LL \chi_{_\lambda} = i \lambda \chi_{_\lambda} ~,~~~ 
\chi_{_\lambda}^* = -\chi_{_{-\lambda}}
\ee
($\lambda$ could be either discrete of continuous), the projection of
$\chi$ on any of the zero modes $\chi_0$
\be
c_0 = \int \frac{d\sigma}{\sqrt{\theta_o}} \, \chi_0^* \, \chi
\ee
commutes with $\chi$ and is a Casimir of (\ref{linP}). The lagrangian
realization $\chi = \LL \varphi$ sets these Casimirs to zero.
It should be stressed, however, that the above Casimirs are an artifact of
the linearized approximation and are lifted in the full Poisson algebra,
since the (omitted) linear in $\chi$ piece endows them with nontrivial
Poisson brackets.

We conclude by pointing out that 
the operator $\omega_o \LL$ generates classical particle motion on the
surface $V = V_o$. So $\varphi$ moves simply as a scalar co-moving
field on this surface. $\chi$, on the other hand, moves as a scalar
density with $\omega_o$ as a weight factor. This is due to the fact that
classical motion on the $V=V_o$ surface also rescales distances away
from the surface, such as $\chi$. $\omega_o \chi$ is a scalar.

\section{Quantization}

The chiral theories obtained in the previous section describe a collection
of fermions in terms of a bosonic field $R$ or $\chi$. This description
is still semiclassical. Quantization of these theories should reproduce
the full quantum mechanics of the Fermi liquid for a large number of
fermions $N$.

This is, indeed, the case for $D=2$, as is well established and we shall
demonstrate below. For
higher dimensions, however, the above chiral bosonic theory will in general
have a continuous infinity of sectors, parametrized by the quotient
manifold $\{\sigma \} / \phi^0$, that is, the manifold of orbits on
the phase space spanned by $\sigma^i$ generated by $\phi^0$. (See remark
2. in section 3.2). Its quantization, therefore, would overcount the degrees
of freedom and would not capture the correct physics of the fermion system.

\subsection{Quantization of the linearized $D=2$ action}

We shall start by studying the most tractable case, which is the
linearized theory in 2 phase space dimensions. This theory would describe
low-lying excitations of a Fermi liquid in one spatial dimension, or edge
excitations in the two-dimensional (integer) quantum Hall state.
The quantization of this theory is fairly standard and will be stated 
here for completeness.

The phase space consists of $\phi^0$ and a unique other variable, call
it $\sigma$. For concreteness, we shall assume that $\sigma$ is periodic
with period $L$. In this case $\LL = \sqrt{\theta_o} \partial_\sigma$ and
$\rho_o = 1/(2\pi \hbar)$. The hamiltonian and Poisson brackets become
\be
\{ \chi (\sigma_1 ) , \chi (\sigma_2 ) \} = 
2\pi\hbar \sqrt{\theta_o (\sigma_+ )} \,
\delta' (\sigma_1 - \sigma_2 )
\label{linP2}
\ee
and
\be
H = \frac{1}{2\pi \hbar} \int_0^L 
\frac{d\sigma}{\sqrt{\theta_o}} \, \half \omega_o \chi^2
\ee
We may change variable from $\sigma$ to a new variable $\tau$ 
defined by $d\sigma = \omega_o \sqrt{\theta_o} d\tau$. This variable
satisfies $\omega_o \LL \tau = 1$; since $\omega_o \LL$ generates classical
particle motion on $R_o$, $\tau$ is the time of flight of a particle
moving on the equipotential line $\phi^o = R_o$. It is also periodic
with a period
\be
T = \int_0^L \frac{d\sigma}{\omega_o \sqrt{\theta_o}}
\ee
We also define a rescaled variable $\alpha = \rho_o \omega_o \chi = 
\omega_o \chi/2\pi\hbar$. 
In terms of $\alpha (\tau )$ the hamiltonian and Poisson brackets become
\be
\{ \alpha(\tau_1 ) , \alpha(\tau_2 ) \} = \frac{1}{2\pi \hbar}
\delta' (\tau_1 - \tau_2 ) ~,~~~
H = 2\pi \hbar \int_0^T d\tau \half \alpha^2
\ee
The above is an abelian chiral algebra and a quadratic hamiltonian.
Quantization can proceed now in a standard way. The quantum commutator
of $\alpha$ is $i\hbar$ times the classical Poisson brackets. In terms of
the Fourier modes
\be
\alpha_n = \int_0^T d\tau \, \alpha(\tau ) e^{-in\frac{2\pi \tau}{T}}
\ee
we obtain the normal-ordered hamiltonian and algebra
\be
[ \alpha_n , \alpha_m ] = n \delta_{n+m} ~,~~~
H = \frac{2\pi \hbar}{T} \sum_{n>0} \alpha_{-n} \alpha_n + 
\frac{\pi \hbar}{T} \alpha_o^2
\ee
The Casimir $\alpha_0$ represents variations of the total particle
number $N$ and will be set to zero, since we work at fixed $N$.
Each pair $(\alpha_{-n} , \alpha_n )$ is a quantum oscillator with
quanta of size $n$, contributing to the energy quanta of size
$\hbar 2\pi n/T$. Calling $b_n = 0,1,2,\dots$ the excitation number
of oscillator $\alpha_n$, the energy eigenvalues are
\be
E = \sum_{n>0} \frac{2\pi \hbar}{T} \, n  b_n
\label{Ebos}
\ee

We may compare this to the states of the fermionic theory. The ground state
consists of a Fermi sea with the first $N$ single-particle levels occupied
by fermions. Low-lying states involve excitations of particles near
the Fermi level, that is, near the $N$-th single-particle energy eigenvalue.

The highly excited energy levels of a particle with hamiltonian $V$ in the
above phase space will be given with good accuracy by the WKB approximation.
The spacing between these levels becomes $\hbar$ times the frequency of
classical periodic motion at this energy, that is, $\hbar 2\pi/T$.
At large $N$ this becomes exact. So the single-particle spectrum becomes
equidistant with the above spacing.

The excitations around the Fermi sea can be parametrized in terms of
the discrete jumps of single fermions to higher single-particle states.
Enumerating fermions in terms of their position in the ground state
with respect to the Fermi level (1 at the top, 2 the one below etc),
their excitations will be $f_n = 0,1,2,\dots$ and the total excitation
energy becomes 
\be
E_f = \sum_n \frac{2\pi \hbar}{T} f_n
\ee
In order to avoid overcounting and respect
Pauli exclusion, we must not allow any particle to jump higher than
the particle above it; that is
\be
f_{n+1} \leq f_n ~,~~~ n=1,2,\dots
\label{ferm}
\ee

We can now establish a mapping between the bosonic and fermionic
states, by parametrizing the $f_n$ as
\be
f_n = \sum_{k \geq n} b_n ~,~~~
b_n = f_n - f_{n+1} ~,~~~ n=1,2,\dots
\ee
This gives a one-to-one correspondence between the unrestricted integers
$b_n$ and the restricted integers $f_n$ satisfying (\ref{ferm}). The
energy of the fermionic system in terms of this parametrization becomes
identical to the bosonic energy (\ref{Ebos}), demonstrating the full
equivalence between the two systems. This is the standard bosonization
picture in terms of phonon excitations of the Fermi sea.

The above analysis in the linearized chiral action shows that it describes
the fermionic system to the leading order in large-$N$. Variations of the
spacing of the levels for finite $N$ will induce $1/N$ corrections.
Further, large excitations depleting the Fermi sea give rise to
nonperturbative effects. The linearized theory misses both corrections.
As we will demonstrate in the next section, the full nonlinear theory
captures $1/N$ corrections but still misses nonperturbative effects.

\subsection{Quantization of nonlinear theories in $D=2$}

The quantization of general nonlinear theories even in $D=2$ is nontrivial.
To make the task tractable, we shall assume that we have identified Darboux
coordinates. So $\theta^{01}=1$, the only dynamical input being the energy
function $V(\phi^0 , \sigma )$.
With an additional canonical transformation we can make the potential to
depend only on the coordinate $\phi^0$.
The coordinate $\phi^1 = \sigma$ will be assumed periodic, reflecting the
fact that $V$=constant lines should be compact so that the system have
discrete levels. Finally, with a residual coordinate transformation
we may make the periodicity of $\sigma$ to be independent of $\phi^0$
and equal to $2\pi$. (Although a canonical transformation could make
$V=\phi^0$, apparently trivializing the problem, this would cause the
conjugate variable $\sigma$ to have a $\phi^0$-dependent periodicity
making the situation unwieldy.) 
The topology of the $\phi_0$ direction will be left unspecified. 

We define the quantum field $\alpha = \rho_o R = R / 2\pi\hbar$,
which is essentially the chiral fermion density in the $\sigma$ direction.
The quantum commutation relations and hamiltonian become
\be
[ \alpha (\sigma_1 ) , \alpha (\sigma_2 ) ] = \frac{i}{2\pi} \partial_\sigma
\delta (\sigma_1 - \sigma_2 ) ~,~~~
H = \frac{1}{2\pi\hbar} \int d\sigma \, U(2\pi \alpha )
\ee
where $U' = V$. $\alpha$ is again a current algebra variable.
Constant and linear terms in $U$ are irrelevant, representing trivial
shifts in energy and a Casimir. Further, a quadratic term in $U$ commutes
with any hamiltonian of the above form, since it corresponds to the
conserved momentum in the $\sigma$ direction, and can be separately
diagonalized. 

A quadratic $U$ is the simplest situation. This would correspond to a
harmonic oscillator potential, with $\phi^0 = \half p^2 + \half x^2$
and $\tan \sigma = x/p$, or to a quantum Hall droplet on a cylinder with
a linear potential along the cylinder axis. (This would, actually,
have two separate boundaries.) Such a $U$ has already been treated
as the linearized case of the previous section. The results, in that
case, are exact to all orders in $1/N$, since the harmonic oscillator
potential has indeed equidistant levels with spacing $\hbar$ times its
classical period.

To analyze a generic potential, we shall first write $\alpha = \alpha_o 
+ {\tilde \alpha}$, where the Casimir $\alpha_o$ is the zero mode of 
$\alpha$ and $\tilde \alpha$ integrates to zero. The particle number is
$N = 2\pi \rho_o R_o = 2\pi \alpha_o$, so $\alpha_o$ is of order $N$, while
we assume that the fluctuating part $\tilde R$ is of order $N^0$.
Writing ${U_o}' = U' (R_o)$ etc., we may expand the hamiltonian as
\be
\frac{1}{2\pi\hbar} \int d\sigma U(2\pi\alpha_o + 2\pi {\tilde R} ) 
= \frac{U_o}{\hbar} + 2\pi\hbar \int d\sigma \bigl[
\frac{{U_o}''}{2} {\tilde \alpha}^2 + 2\pi\hbar 
\frac{{U_o}'''}{6} {\tilde \alpha}^3 + \dots \bigr]
\ee
Each subsequent term in the expansion is subleading in $\hbar$ and $1/N$.
The leading term is the quadratic
one, which has already been treated and commutes with all subsequent
terms. So we may omit it and examine only the subleading corrections
to this hamiltonian. We see that the first nontrivial term, of order
$1/N$ is the cubic one.

A pure cubic $U$ would correspond to free particles on a circle, with
$\phi^0 = p$ and $\sigma = x$. The cubic hamiltonian with two chiral 
fields $R$ (representing the two Fermi levels) is also known as the 
fermionic collective field theory \cite{JeSa},\cite{Pol}.
It is well established that the cubic bosonic hamiltonian reproduces
{\it exactly} the states of the fermionic system to all orders in the
$1/N$ expansion (which stops at $O(1/N)$ in this case) \cite{JeSa}. 

We conclude that the nonlinear hamiltonian for any function $U(R)$
reproduces the correct quantum mechanical results at least up to order
$1/N$. For higher orders we should also include derivative terms,
which would arise from renormalization and start at order $1/N^2$.
Nonperturbative effects connected to the depletion of the Fermi sea will 
cause corrections of order $e^{-O(N)}$ and will always be missed.

\subsection{Anyons, Calogero, FQHE states}

In all previous considerations we took the density of the droplet to
be $\rho_o = 1/2\pi\hbar$. Although this is sensible from the fermion
point of view, corresponding to one fermion per quantum state, it is
not really required. Both the classical and the quantum theory are
well-defined for any $\rho_o$.

We may, thus, choose to work with $\rho_o = \nu /2\pi\hbar$, where
$\nu$ represents the fraction of the single-particle states occupied
by particles and can take any value. We will obtain a semiclassical
description of particles with generalized exclusion statistics 
where each particle occupies $\ell=1/\nu$ states. For $D=2$ such a system 
would describe either interacting Calogero-Sutherland particles with
strength of the inverse-square two-body potential $\ell (\ell-1)$ 
\cite{PolC},\cite{Isa}, or anyons in the lowest Landau level with
statistical exchange phase $\alpha_s = \pi \ell$ \cite{LeMy},
or Laughlin states in the fractional quantum Hall effect (FQHE)
with exponent $\ell$ \cite{Lau}.

Quantization of this system along the lines of the previous section
in terms of the density field $\alpha = \rho_o R = R/2\pi\hbar\ell$
leads to the algebra and hamiltonian
\be
[ \alpha (\sigma_1 ) , \alpha (\sigma_2 ) ] = \frac{i}{2\pi \ell} 
\partial_\sigma \delta (\sigma_1 - \sigma_2 )
\label{Cell}
\ee
\be
H = \frac{1}{2\pi\hbar\ell} \int d\sigma U ( 2\pi \hbar\ell \alpha )
\label{Hell}
\ee
For a quadratic $U$ corresponding to a linear potential $V(R)$, which
again describes particles in a harmonic oscillator potential, or a linear
potential on the cylinder for the lowest Landau level, the results
are identical to the free fermion case ($\ell =1$), as can be seen by
working with the rescaled field ${\sqrt \ell} \alpha$ that eliminates
$\ell$ from both (\ref{Cell}) and (\ref{Hell}). 
Indeed, it is known that the energy eigenstates of the Calogero model
\cite{Cal}, or of anyons in the lowest Landau level 
\cite{DuLeTru}-\cite{GruHaKaWe}, or of Laughlin particles, in a
harmonic potential are identical to those of fermions up to a ground
state shift. The density and particle correlation functions, on the other
hand, constructed in terms of $\alpha$ or its exponential, will exhibit
nontrivial scaling due to the appearance of $\ell$ in (\ref{Cell}).

For a cubic $U$, corresponding to free particles on a periodic domain,
the above will give the leading $1/N$ terms of the collective field
description of the Calogero model and will reproduce the spectrum to
leading order in $1/N$ \cite{AnBa},\cite{MiPo}.
We see that the exact results obtained for 
fermions ($\ell=1$) in the cubic case do not carry over to the 
Calogero/anyon/FQHE case. In the limit of large $N$ and fixed gap for
the low-lying states (the conformal field theory limit), at any rate,
the above theory reproduces the correct spectrum and correlations.

\section{Conclusions and discussion}

We have presented an analysis of the phase space dynamics of droplets
in an arbitrary phase space
and derived their hamiltonian and canonical structure. In the linearized
approximation we recovered the chiral actions of edge excitations.
The nonlinear hamiltonian classically describes arbitrary deformations
of the droplet and quantum mechanically captures $1/N$ corrections,
at least in the two dimensional case.

The transition from the hamiltonian to an action in the full nonlinear
case is still an open issue, due to the nontrivial form of the Poisson
brackets (\ref{PR}). If we work with Darboux coordinates, and choose
one of them for $R$, the algebra becomes affine. The advantage of the
linearized case is that the right hand side of (\ref{PR}) becomes
a constant and we have essentially a generalized Heisenberg algebra
that can be easily realized with a lagrangian.

The theories obtained here are first-order and chiral, in the same sense
that propagation on a phase space is chiral since from each point it proceeds
to a unique direction. They represent a phase space bosonization of the
fermionic systems they describe. This bosonization, however, fails quantum
mechanically. First, there is the possibility that the quantum action
will get renormalized. This is not fatal and we have already commented
on it in the last section. Secondly, and more importantly, in dimensions
higher than two the linearized effective action describes chiral propagation
along the direction of classical motion, while the remaining of the
phase coordinates are inert. They act thus as parameters in the theory,
giving rise to an infinity of modes which will in general overcount the
quantum mechanical modes. Nontrivial topology of the flow lines on the
surface of the droplet may connect and reduce some of these modes, but
it is unlikely that we will get the correct description.

On a more speculative level, the fractional quantum Hall state
has been formulated in terms of a noncommutative Chern-Simons
field theory \cite{Sus}, and its boundary excitations can be described as a
matrix model \cite{PolQ}. These descriptions capture essential quantum
properties of the state, such as the discrete nature of the underlying
constituent electrons and their statistical repulsion, at the semiclassical
level. It would be interesting to examine if these (matrix) features can be
incorporated in the phase space description of fermions or, in general,
particles with mutual exclusion.

Finally, we only dealt with abelian theories. The nonabelian case
requires the formulation of phase space droplets for several flavors
or spin states of fermions. This and other related matters will be dealt with
in a different publication.
\vskip 0.2in

{\it \underline {Acknowledgements}:} I am thankful to Roman Jackiw,
Hans Hansson, Dimitra Karabali, Parameswaran Nair and David Schmeltzer
for useful comments on the manuscript.
This research was supported in part by the
National Science Foundation under grant PHY-0353301 and by the
CUNY Research Foundation under grant PSC-CUNY-66565-0035.

\end{document}